\documentclass[manuscript]{aastex}
\usepackage{emulateapj5}
\usepackage{graphicx}

\newcommand{\La}{\mbox{${\rm Ly\alpha}$}}

\newcommand{\Ha}{\mbox{${\rm H\alpha}$}}

\newcommand{\Line}[3]{\Ion{#1}{#2}\,$\lambda$\,#3}
\newcommand{\Lines}[3]{\Ion{#1}{#2}\,$\lambda\lambda$\,#3}
\newcommand{\Ion}[2]{#1{\,\scriptsize #2}}
\newcommand{\Twd}{\mbox{$T_{\rm wd}$}}

\newcommand{\Porb}{\mbox{$P_{\rm orb}$}}

\newcommand{\Msun}{\mbox{$M_{\odot}$}}

\newcommand{\ecs}{\mbox{$\rm erg\;cm^{-2}s^{-1}$}}

\begin{document}

\title{Anomalous ultraviolet line flux ratios in the cataclysmic
variables 1RXS\,J232953.9+062814,  CE315,  BZ\,UMa and EY\,Cyg  observed
with \textit{HST}/STIS \altaffilmark{1}}

\author{Boris T. G\"ansicke}
\affil{Department of Physics and Astronomy, University of Southampton,
Highfield, Southampton SO17 1BJ, UK}
\email{btg@astro.soton.ac.uk}

\author{Paula Szkody}
\affil{Astronomy Department, University of Washington, Seattle, WA 98195}
\email{szkody@astro.washington.edu}
\author{Domitilla de Martino}
\affil{Osservatorio di Capodimonte, Via Moiariello 16, I-80131 Napoli, Italy}
\email{demartino@astrna.na.astro.it}
\author{Klaus Beuermann}
\affil{Universit\"ats-Sternwarte, Geismarlandstr. 11, 37083 G\"ottingen, 
Germany}
\email{beuermann@uni-sw.gwdg.de}
\author{Knox S. Long}
\affil{Space Telescope Science Institute, 3700 San Martin Drive,
Baltimore, MD 21218;
long@stsci.edu}
\author{Edward M. Sion}
\affil{Department of Astronomy and Astrophysics, Villanova 
University, Villanova, PA 19085}
\email{emsion@ast.villanova.edu}
\author{Christian Knigge, Tom Marsh}
\affil{Department of Physics and Astronomy, University of Southampton,
Southampton SO17 1BJ UK}
\email{christian@astro.soton.ac.uk, trm@astro.soton.ac.uk}
\author{Ivan Hubeny}
\affil{Laboratory for Astronomy and Solar Physics, NASA/GSFC,
Greenbelt, MD 20711} 
\email{hubeny@tlusty.gsfc.nasa.gov}


\altaffiltext{1}{Based on observations made with the NASA/ESA Hubble Space
Telescope, obtained at the Space Telescope Science Institute, which is
operated by the Association of Universities for Research in Astronomy,
Inc., under NASA contract NAS 5-26555.}

\begin{abstract}
Brief \textit{HST}/STIS spectroscopic snapshot exposures of the
cataclysmic variables 1RXS\,J232953.9+062814, CE315, BZ\,UMa and EY\,Cyg 
reveal very large \Ion{N}{V}/\Ion{C}{IV} line flux ratios,
similar to those observed in AE\,Aqr. Such anomalous line flux ratios
have so far been observed in 10 systems, and presumably reflect a
different composition of the accreted material compared to the
majority of cataclysmic variables.  We discuss the properties of this
small sample in the context of the recent proposal by
\citet{schenkeretal02-1} that a significant fraction of the
present-day population of cataclysmic variables may have passed
through a phase of thermal time-scale mass transfer.
\end{abstract}

\section{Introduction}
Spectroscopic studies of cataclysmic variables (CVs) are exceptionally
rewarding in the far-ultraviolet (FUV), as this wavelength range contains
transitions of diagnostically important metals (e.g. C, N, O, Si) in
various ionization stages that are not accessible from the ground.
Analyzing the observed ultraviolet lines provides information on both
the ionizing spectrum emitted by the hot accretion region on/near the
white dwarf, as well as on the abundances of the material in the
accretion flow. Unfortunately, high-quality ultraviolet spectroscopy
is rather difficult to obtain, and all large-scale statistical studies have so
far been based on data obtained with \textit{IUE}
\citep[e.g.][]{verbunt87-1, ladous91-1, demartino95-1, maucheetal97-1}.

We have initiated a FUV spectroscopic survey of CVs using
the Space Telescope Imaging Spectrograph (STIS) on-board the
\textit{Hubble Space Telescope} (\textit{HST}). The observations are
carried out as \textit{snapshots}, gaps in the \textit{HST} schedule
that can not be filled with regular observations. This survey will
eventually produce a homogenous database of modern high-quality
ultraviolet spectra for a large number of CVs.
Here we report the first results obtained from this program, namely
the measurement of an extreme \Ion{N}{V}/\Ion{C}{IV} emission line
flux ratio in 1RXS\,J232953.9+062814 (henceforth RX\,J2329), a
recently discovered dwarf nova with an unusually short orbital period
of 66\,min \citep{thorstensenetal02-1}, and in CE315, an AM\,CVn type CV
\citep{ruizetal01-1}. In addition, we confirm the unusually large
\Ion{N}{V}/\Ion{C}{IV} emission line flux ratios in the dwarf novae
BZ\,UMa and EY\,Cyg, first detected in \textit{IUE} data by
\citet{winter+sion01-1} and \citet{sion02-1}.

\section{FUV spectroscopy}
\begin{figure*}
\includegraphics[angle=-90,width=8cm]{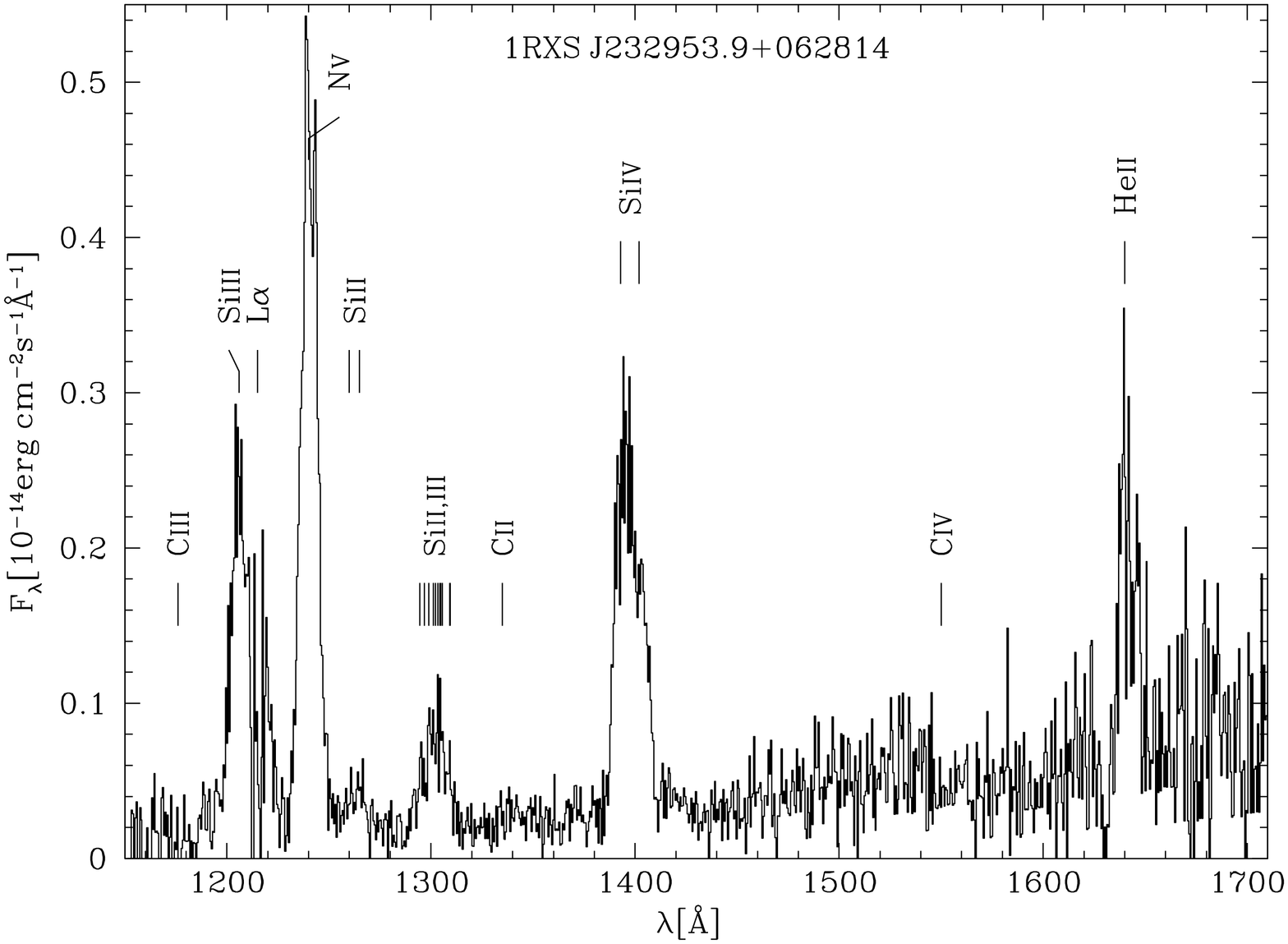}
\includegraphics[angle=-90,width=8cm]{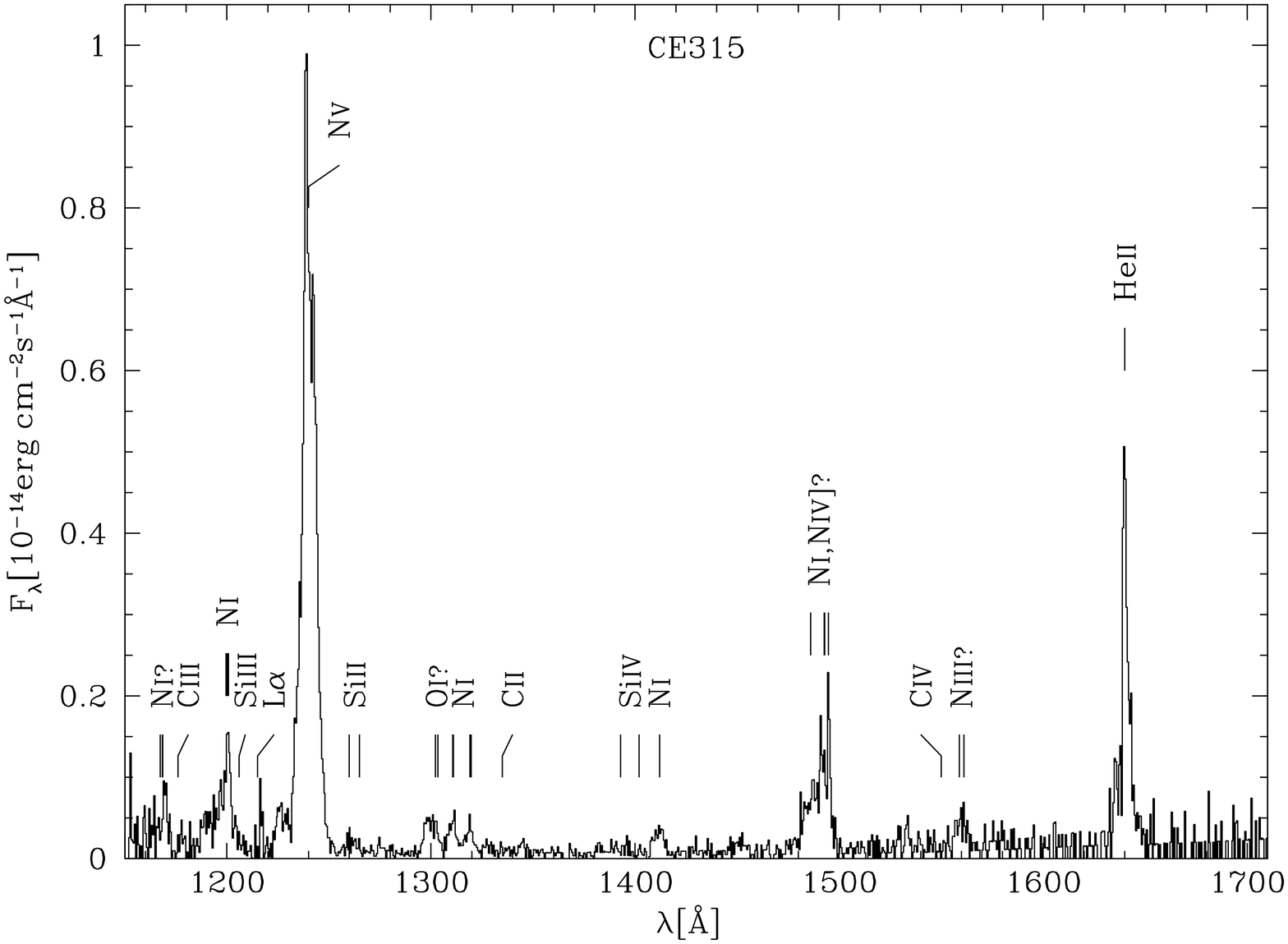}\\
\includegraphics[angle=-90,width=8cm]{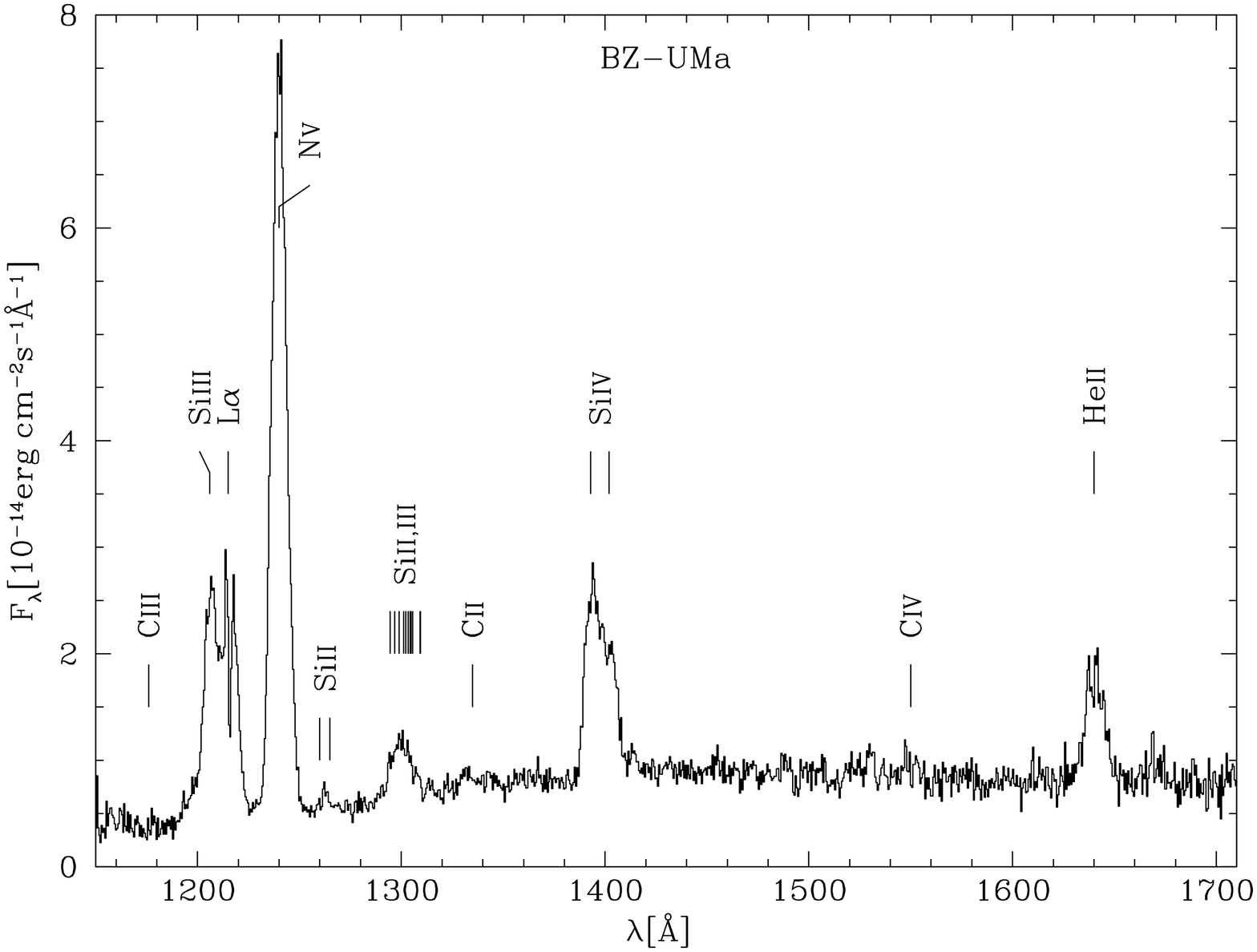}
\includegraphics[angle=-90,width=8cm]{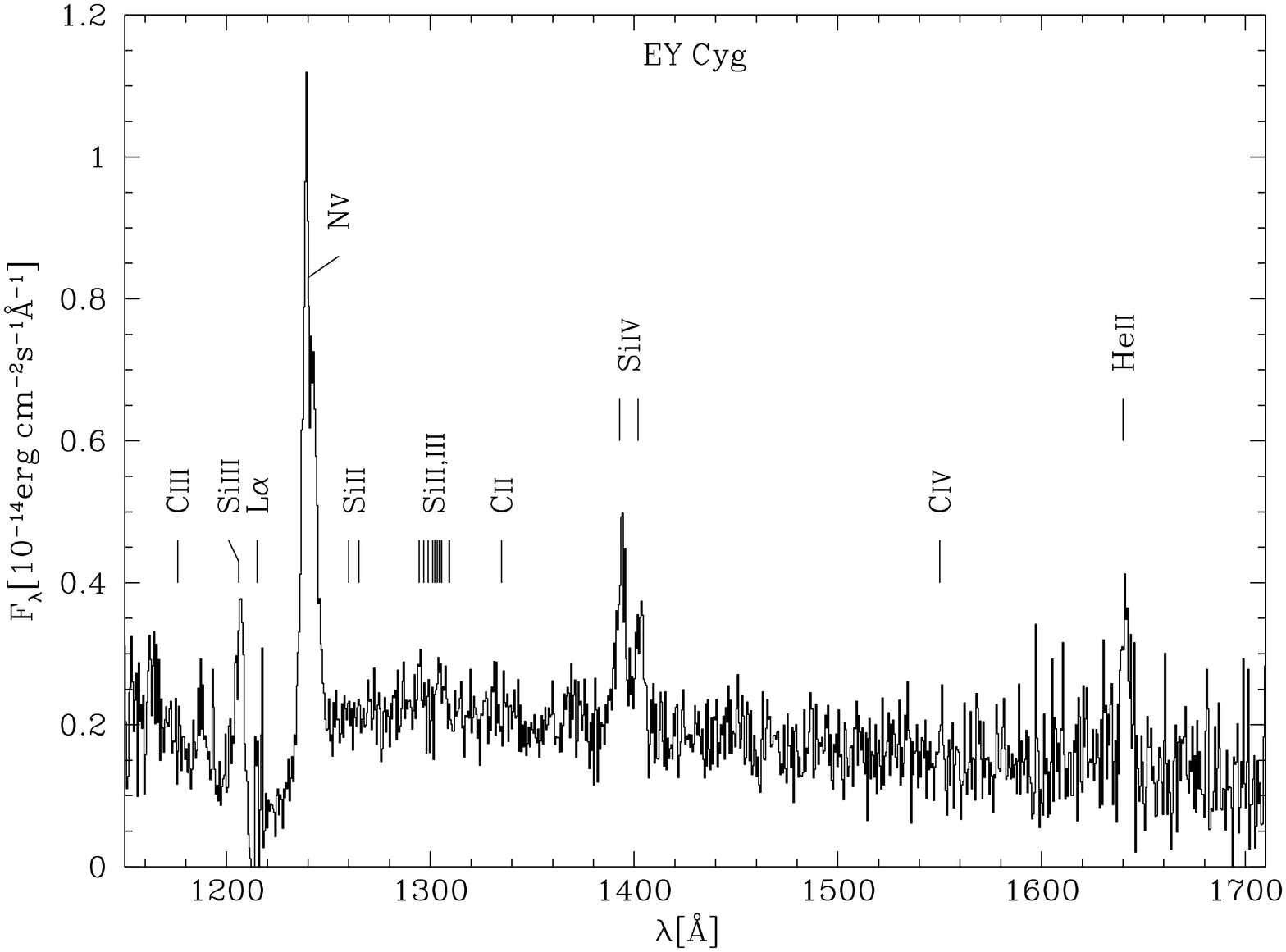}
\caption{\label{f-stisspectra}\textit{HST}/STIS G140L snapshot spectra
of RX\,J2329, CE315, BZ\,UMa, and EY\,Cyg.}
\end{figure*}

\textit{HST}/STIS FUV spectroscopy of RX\,J2329, CE315, BZ\,UMa
and EY\,Cyg was obtained using the G140L grating and the
$52\arcsec\times0.2\arcsec$ aperture (Table\,\ref{t-hstobs}). This
instrumental setup provides a spectral resolution of
$R\approx1000$. The data were processed with the latest release of
CALSTIS (V2.13b), which takes into account the decaying sensitivity of
the G140L grating. The STIS spectra of RX\,J2329, CE315, BZ\,UMa and EY\,Cyg
are displayed in Fig.\,\ref{f-stisspectra}. An apparent similarity
between all four systems is the unusually large strength of
\Line{N}{V}{1240}, as well as the lack of noticeable
\Line{C}{III}{1176}, \Line{C}{II}{1335} or \Line{C}{IV}{1550} emission.

Despite the short exposure times, the quality of our STIS spectra
exceeds that of the previous \textit{IUE} observations of BZ\,UMa and
EY\,Cyg by large factors.  The continuum FUV spectrum of BZ\,UMa is
typical for a short-period dwarf nova, being most likely a mixture of
emission from the accretion disc and the white dwarf. The fact that
the broad \La\ absorption originating in the white dwarf photosphere
is largely filled in with emission suggests that the disc contributes
significantly in the FUV. In contrast to this, the FUV
continuum of RX\,J2329 is extremely red, suggesting that the white
dwarf in this system is very cold (reddening is negligible for the
distance of $\sim200$\,pc determined by
\citealt{thorstensenetal02-1}).  The continuum slope and \La\
absorption observed in EY\,Cyg is reminiscent of the photospheric
emission of a white dwarf with $\Twd\sim25\,000$\,K.  A detailed
modelling of the continuum spectra of RX\,J2329, BZ\,UMa and EY\,Cyg,
will be presented along with the data of additional objects from our
snapshot survey in a separate paper.

The STIS spectrum of CE315 differs from that of the other three
systems in that practically no continuum emission is
detected. Furthermore, none of the usual emission lines of silicon
are present in the spectrum. Most of the weak emission lines are likely
due to \Ion{N}{I}. The identifications of emission features at
$\sim1170$\AA, $\sim1300$\,\AA, $\sim1490$\,\AA\ and $\sim1560$\AA\
remain somewhat uncertain. With respect to the last line, we stress,
however, that this feature is clearly not related to emission of
\Ion{C}{IV}.

\section{Emission Line Flux Ratios}
We have measured the emission line fluxes of \Line{N}{V}{1240},
\Lines{Si}{IV}{1394,1403}, \Line{C}{IV}{1550}, and \Line{He}{II}{1640}
in RX\,J2329, CE315, BZ\,UMa and EY\,Cyg adopting the following two
methods: (1) We fitted a small interval around the central wavelength
with a linear function plus a Gaussian line (in the case of
\Ion{Si}{IV} two Gaussians), and computed the line flux from the
Gaussian fit parameters; (2) we used the \texttt{integrate/line}
command in the \texttt{ESO/MIDAS} suite to interactively determine the
continuum level and integrate the flux of the emission line(s) above
this continuum. The results from these two methods were found to be in
good agreement, and the average of the line fluxes from both
measurements are reported in Table\,\ref{t-linefluxes}. 
Note that we did not detect \Line{C}{IV}{1550} at a
statistically significant level in any of the four stars. In addition,
\Line{Si}{IV}{1394,1403} were not detected in CE315.  In order to
constrain the \Ion{C}{IV} and \Ion{Si}{IV} line fluxes we have added to
the STIS spectra Gaussian profiles of similar width as the observed
lines and analysed the synthetic emission lines as described above. We
repeated this process, decreasing the line fluxes of the added
Gaussian lines, until reaching the detection threshold. The upper
limits on the line fluxes derived in that way are given in
Table\,\ref{t-linefluxes}.

For completeness, we compare the \Ion{N}{V} emission line fluxes
obtained from our STIS spectra of BZ\,UMa and EY\,Cyg with the values
derived from archival \textit{IUE} observations (the other lines are
too weak in the \textit{IUE} spectra to permit any useful
measurement).  In the case of BZ\,UMa, two SWP spectra are available
(32778, 32783), which contain \Ion{N}{V} at a similar flux level as
our STIS spectrum ($\sim60\pm5\times10^{-13}$\,\ecs). In the case of
EY\,Cyg, the two availabe SWP spectra that have a sufficient
signal-to-noise ratio suggest that in this system the \Ion{N}{V}
line may vary significantly (33428: $20\pm4\times10^{-14}$\,\ecs;
31304: $10\pm3\times10^{-14}$\,\ecs). Second-epoch STIS observations
of EY\,Cyg would be desirable to confirm the variability.

\section{Discussion}
To our knowledge, 10 CVs have so far been reported to display
anomalously high \Ion{N}{V}/\Ion{C}{IV} emission line flux ratios:
AE\,Aqr \citep{jamesonetal80-1}, BY\,Cam
\citep{bonnet-bidaud+mouchet87-1}, TX\,Col \citep{mouchetetal91-1},
V1309\,Ori \citep{szkody+silber96-1}, MN\,Hya
\citep{schmidt+stockman01-1}, EY\,Cyg
(\citealt{winter+sion01-1,sion02-1}), BZ\,UMa
(\citealt{winter+sion01-1,sion02-1}), GP\,Com
\citep{lambert+slovak81-1, marshetal95-2}, RX\,J2329, and CE315.  We
show in Figure\,\ref{f-lineratios} the line flux ratios
\Ion{N}{V}/\Ion{C}{IV} vs. \Ion{Si}{VI}/\Ion{C}{IV} and
\Ion{He}{II}/\Ion{C}{IV} vs. \Ion{Si}{IV}/\Ion{C}{IV} of these
``anomalous'' CVs, except for CE315 (in which neither \Ion{C}{IV} nor
\Ion{Si}{IV} are detected, Fig.\,\ref{f-stisspectra}) and GP\,Com (in
which \Ion{Si}{IV} is not detected, and \Ion{He}{II} is not included
in the wavelength range covered by the \textit{HST}/FOS
observations). The line flux measurements are taken from the
\textit{IUE} study of \citet{maucheetal97-1}, except for V1309\,Ori
\citep{szkody+silber96-1,schmidt+stockman01-1}, MN\,Hya
\citep{schmidt+stockman01-1}, TX\,Col (\citealt{mouchetetal91-1}), the
line ratios were measured as described above from the \textit{IUE}
spectra SWP00502L and SWP34129L), and RX\,J2329, BZ\,UMa, and EY\,Cyg
(Table\,\ref{t-linefluxes}).

\begin{figure*}
\includegraphics[width=8cm]{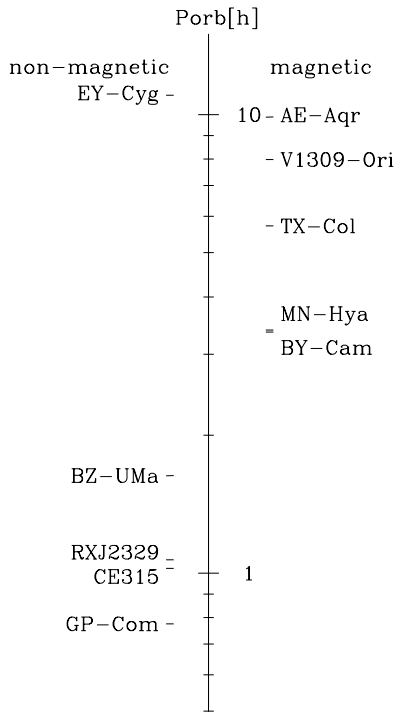}
\caption{\label{f-anomalous}
Properties of the seven CVs known to display significantly enhanced
\Ion{N}{V}/\Ion{C}{IV} line flux ratios.}
\end{figure*}

The largest \Ion{N}{V}/\Ion{C}{IV} (Fig.\,\ref{f-lineratios},
Table\,\ref{t-linefluxes}) ratios are observed in AE\,Aqr during
quiescence, RX\,J2329, CE315, BZ\,UMa and EY\,Cyg, followed by
BY\,Cam, V1309\,Ori, MN\,Hya and GP\,Com and finally by TX\,Col with
a moderate enhancement of this line flux ratio with respect to
``normal'' CVs. Figure\,\ref{f-anomalous} shows the period
distribution of the ``anomalous'' systems, and indicates whether the
accreting white dwarf possesses a significant magnetic field or not.

A number of possible explanations for the observed anomalous line
ratios have been proposed. After the first \textit{IUE} detection of
an inverse \Ion{N}{V}/\Ion{C}{IV} emission line flux ratio in BY\,Cam,
\citet{bonnet-bidaud+mouchet87-1} discussed peculiar ionization
conditions of the emitting region, and the possibility of anomalous
abundances in the accreting material, either as the consequence of a
(moderately recent) nova outburst or related to the nuclear evolution
of the donor star. \citet{maucheetal97-1} have cast some doubt on the
hypothesis of anomalous abundances, and suggested that confusion of
\Lines{N}{V}{1238.8,1242.8} doublet with
\Lines{Mg}{II}{1239.9,1240.4}, combined with a low ionization
parameter could explain the observed line ratios. However,
\citet{maucheetal97-1} also noted that none of their photoionization
models was able to reproduce the strong correlation between the
observed \Ion{N}{V}/\Ion{C}{IV} and \Ion{Si}{IV}/\Ion{C}{IV} line flux
ratios.  

The systems displaying the most ``anomalous'' line flux ratios span a
very large range in orbital periods (Fig.\,\ref{f-anomalous}),
corresponding to a  large range in binary sizes and accretion
rates. In addition, the white dwarfs contained in these systems cover
a wide range in magnetic field strength. Considering this large degree
of variety, it appears rather coincidental that the ionization
conditions in all those systems could be similar enough to favour the
observed extreme line ratios.  We note in passing that an extreme
\Ion{N}{V}/\Ion{C}{IV} line flux ratio has also been observed in the
$\Porb=4.1$\,h soft X-ray transient XTE\,J1118+480
\citep{haswelletal02-1}. This finding extends the parameter range over
which such anomalous line ratios have been observed in terms of the
ionizing X-ray spectrum and luminosity emitted by the accreting
compact object.  Finally, \citet{mouchetetal03-1} recently detected an
unusually large \Ion{N}{III}/\Ion{C}{III} line flux ratio in
\textit{FUSE} observations of BY\,Cam. Their photoionization
models strongly favour non-solar abundances over a peculiar ionizing
spectrum. We will, therefore, assume in the following discussion that
the differences in the observed line ratios reflect differences in the
abundances of the material accreted from the donor star.  Two
different processes can result in non-solar abundances in the envelope
of the secondary.

(1) The companion star may be contaminated by CNO processed material
from the nova shell ejected by the white dwarf during a classical nova
explosion.  Unfortunately, no detailed predictions on the relative
nitrogen and carbon abundances are available for such ``polluted''
companion stars in recent post-novae (\citealt{stehle+ritter99-1,
  marksetal97-1, marks+sarna98-1}, see also the discussion by
\citealt{sionetal98-1}) impeding a quantitative comparison with the
observations.

(2) A second possibility to achieve mass transfer of material with
highly enhanced N/C abundance ratios is that the CV originally
contained a donor \textit{more massive} than the white dwarf,
resulting in a short phase of (unstable) thermal time-scale mass
transfer (TTSMT). Recent calculations by \citet[][ see also
  \citealt{podsiadlowskietal01-1}]{schenkeretal02-1} show that systems
with an initial donor mass of up to 2\,\Msun\ may survive the TTSMT,
and evolve thereafter into an (apparently) normal CV. However, in such
systems, the white dwarf accretes from the CNO processed core of the
previously more massive donor, stripped of its outer layers during the
TTSMT. The high accretion rate during the TTSMT can result in stable
hydrogen shell burning on the white dwarfs. It is likely that some (if
not all) of the known supersoft X-ray binaries are the observational
counterparts to the TTSMT CVs containing a shell-burning white dwarf
\citep{kahabka+vandenheuvel97-1,
  gaensickeetal00-3}. \citet{schenkeretal02-1} make a number of
predictions for the population of CVs that have gone through a phase
of TTSMT which we confront here with the observed properties of the
CVs displaying an anomalous \Ion{N}{V}/\Ion{C}{IV} line flux ratio.

\textit{AM\,CVn and V485\,Cen stars as post-SSXBs:}
\citet{schenkeretal02-1} and \citet{podsiadlowskietal01-1} show that
CVs going through a phase of TTSMT have a good chance ending up as
AM\,CVn stars with very short orbital periods and degenerate donor
stars. GP\,Com and CE315 are the only two known AM\,CVn stars which
display strong FUV emission lines, and both systems show strongly
enhanced \Ion{N}{V}/\Ion{C}{IV} line flux ratios, as expected for
post-TTMST CVs (we note, however, that also the alternative evolution
channel for AM\,CVn, via double degenerate binaries, 
predicts enhanced N/C ratios).  Two CVs are known to be intermediate
objects between the hydrogen-deficient AM\,CVn stars and the
``normal'' hydrogen-rich ($\Porb\ga80$\,min) CVs: V485\,Cen
($\Porb=59$\,min) and RX\,J2329
($\Porb=64$\,min). \citet{schenkeretal02-1} predict these systems to
be descended from SSXBs. The high \Ion{N}{V}/\Ion{C}{IV} flux ratio
observed in RX\,J2329 lends support to this suggestion.

\textit{Are the spectral types of the ``anomalous line ratio'' CVs
``anomalous''?} \citet{schenkeretal02-1} predict that the donor stars
in CVs which went through a phase of TTSMT should have spectral types
too late for their orbital period. This is indeed the case for AE\,Aqr
and V1309\,Ori.  The M$5.5\pm0.5$V and M$3-4$V donors in BZ\,UMa
\citep{ringwaldetal94-1} and MN\,Hya \citep{ramsay+wheatley98-1},
respectively, are, however, rather ``normal'' for the orbital periods
of the two systems.  The donor in RX\,J2329 is much \textit{too early}
for the 64\,min period \citep{thorstensenetal02-1}~--~but it has the
excuse that the concept of ``normal'' spectral type does not apply to
this period range\footnote{\citet{thorstensenetal02-2} have
recently identified a second short-period dwarf nova, QZ\,Ser, which
contains a secondary with a spectral type too early for its orbital
period. The authors detected enhanced sodium absorption in the optical
spectrum of QZ\,Ser and argue that the donor star in this system may have
undergone significant nuclear evolution.}
\citet{tovmassianetal02-1} recently determined the
spectral type of the donor in EY\,Cyg to be K0, but with a radius 1.6
times larger than a main sequence star. Interestingly,
\citet{tovmassianetal02-1} suspected that EY\,Cyg has recently gone
through a classical nova explosion, however, their \Ha\ images of the
system do not unambiguously reveal signs of a nova shell. The spectral
type of the donor star in TX\,Col is only poorly constrained, and
nothing is known about the donor in BY\,Cam.

\textit{How many CVs display a N/C enhancement?}
\citet{schenkeretal02-1} suggest that a large fraction, up to one
third, of the CV population might be descended from supersoft X-ray
binaries. The sample of CVs analyzed by \citet{maucheetal97-1}
contained data for 20 systems, of which two (BY\,Cam and AE\,Aqr)
display unusually large \Ion{N}{V}/\Ion{C}{IV} line ratios. At the
time of writing, we have obtained \textit{HST}/STIS snapshot spectra
of 31 CVs with strong emission lines, and found four systems with
extremely enhanced \Ion{N}{V}/\Ion{C}{IV} ratios. While the presently
available sample is still too small for any definite conclusion, it
suggests that the fraction of CVs that have gone through a phase of
thermal time-scale mass transfer and that display large abundance
anomalies could well be of the order of $10-15$\,\%.

While Fig.\,\ref{f-anomalous} may suggest that the occurence of
``anomalous'' CVs is higher among magnetic systems, we stress that
such a conclusion has to be treated with great care because of the
involved selection effects. The study of \citet{maucheetal97-1}
contains 7 magnetic CVs in a total sample of 20 systems, which
suggests that the fraction of magnetic CVs with sufficiently good
\textit{IUE} observations is larger than that of non-magnetic CVs.
Our STIS sample contains 5 magnetic CVs out of 31 emission-line
systems, and all systems displaying large \Ion{N}{V}/\Ion{C}{IV} line
flux ratios within this sample are non-magnetic. 

As a final note, we recall that evidence for carbon-depletion and
nitrogen-enhancement has also been found in FUV observations of the
accreting white dwarfs in several dwarf novae (e.g. U\,Gem:
\citealt{sionetal98-1, long+gilliland99-1} and VW\,Hyi:
\citealt{sionetal01-2}), and have so far been discussed primarily in
the context of the aftermath of a (recent) nova explosion.

\section{Summary}
Our \textit{HST}/STIS snapshot spectra of RX\,J2329, CE315, BZ\,UMa,
and EY\,Cyg reveal extremely large \Ion{N}{V}/\Ion{C}{IV} and
\Ion{Si}{IV}/\Ion{C}{IV} line flux ratios, similar to those observed
in AE\,Aqr. Such anomalous line flux ratios are expected in CVs that
went through a phase of thermal timescale mass transfer and now
accrete CNO processed material from a companion stripped of its
external layers. A full understanding of the possible links between
the observed anomalous line ratios and the evolution of CVs requires a
significantly larger sample of systems for which high-quality
ultraviolet spectroscopy is available. Our Cycle\,11 snapshot program
and its approved continuation in Cycle\,12 will eventually provide
such a data base.

\begin{figure*}
\includegraphics[angle=-90,width=8cm]{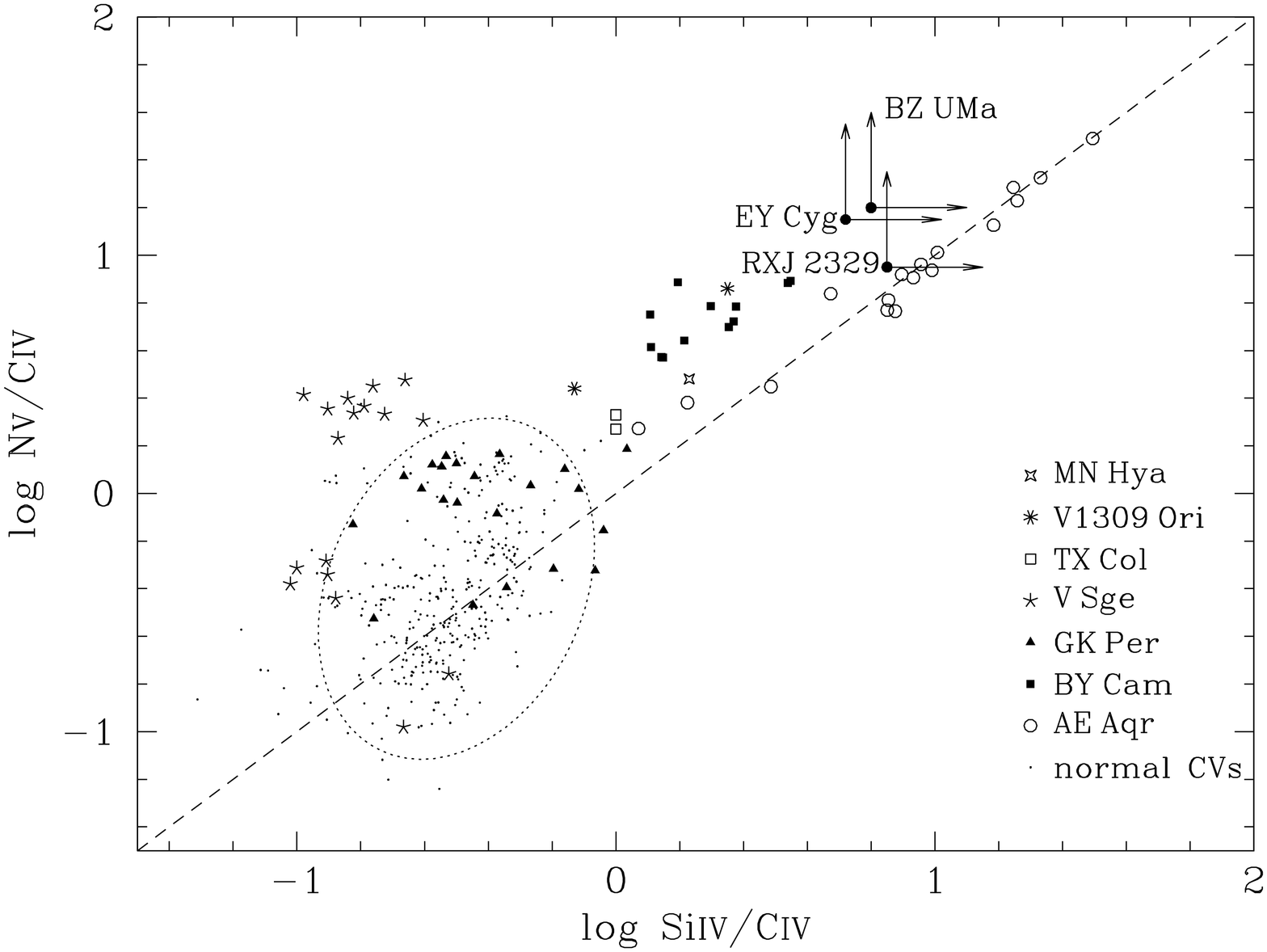}
\includegraphics[angle=-90,width=8cm]{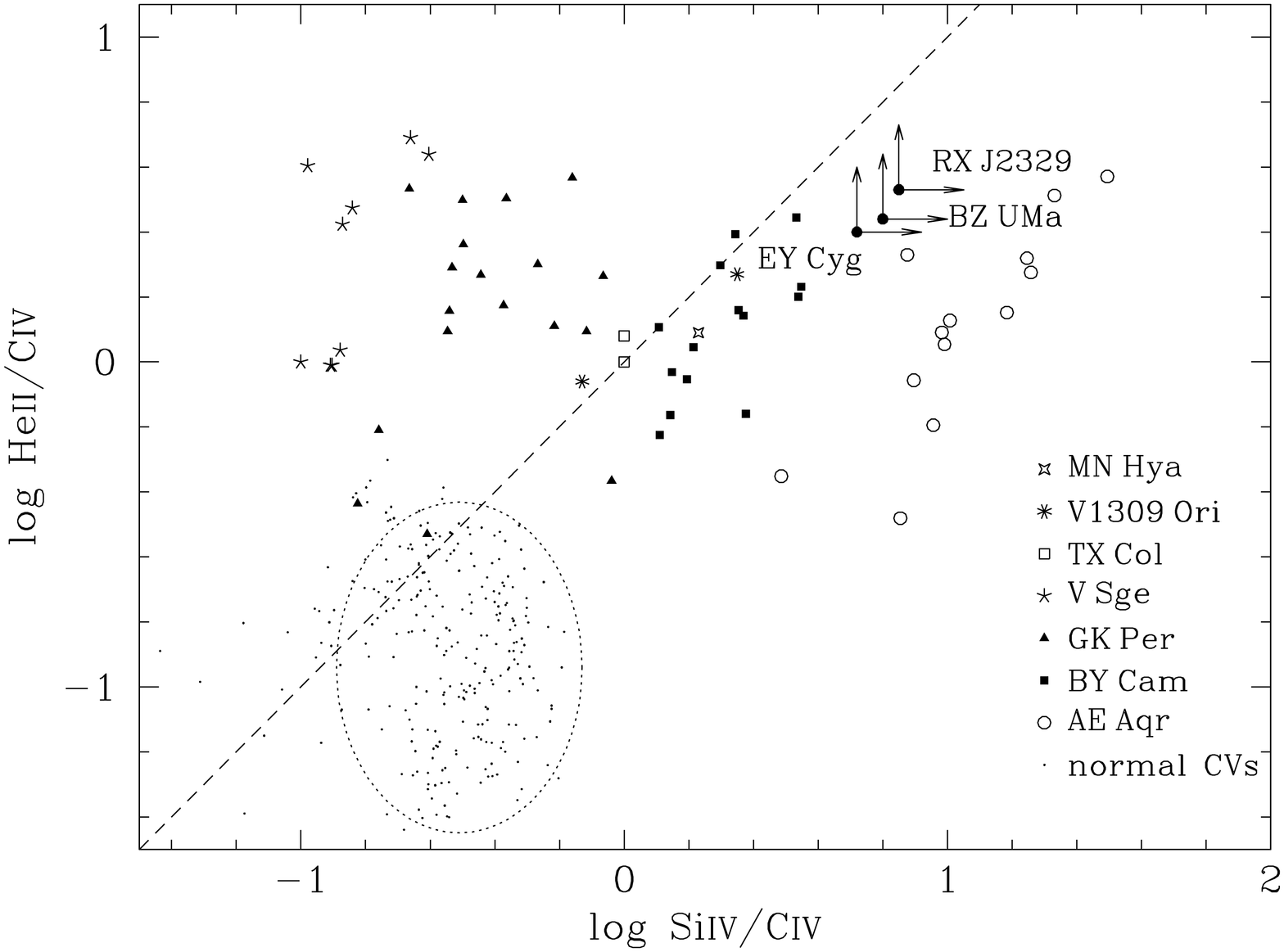}
\caption{\label{f-lineratios} Far ultraviolet line flux ratios in
CVs. Most ``normal'' CVs, in which the white dwarf presumably accretes
from a donor star with (nearly) solar-abundances, fall within a
limited parameter space (roughly indicated by the dotted ellipses). The
``anomalous'' line flux ratios observed in BZ\,UMa, EY\,Cyg,
RX\,J2329, AE\,Aqr, BY\,Cam, V1309\,Ori, MN\,Hya, and TX\,Col suggest
that the white dwarfs in these systems accrete material with an
enhanced N/C abundance ratio. The line flux ratios of RX\,J2329,
BZ\,UMa and EY\,Cyg have been measured from our \textit{HST}/STIS
snapshot spectra. The line flux ratios of V1309\,Ori and MN\,Hya are
taken from \citet{szkody+silber96-1} and \citet{schmidt+stockman01-1},
that of TX\,Col was obtained from archival \textit{IUE} data.  All
other line ratio data are based on \textit{IUE} observations and were
kindly supplied by C. Mauche, see Fig.\,8 in \citet{maucheetal97-1}.}
\end{figure*}

\acknowledgements{We acknowledge C. Mauche for kindly supplying us
with the IUE line ratio data shown in Fig.\,\ref{f-lineratios}, and
S. Araujo-Betancor for a careful reading of the
manuscript. The referee, John Thorstensen, is thanked for
providing a timely report and some useful suggestions. BTG was
supported by a PPARC Advanced Fellowship.  Additional support was
provided through NASA grant GO-9357 from the Space Telescope Science
Institute, which is operated by the Association of Universities for
Research in Astronomy, Inc., under NASA contract NAS 5-26555.}

\bibliographystyle{apj}

\newpage

\begin{deluxetable}{lccc}
\tablecolumns{5}  
\tablewidth{0pc}  
\tablecaption{\label{t-hstobs}Log of the \textit{HST} observations}
\tablehead{  
\colhead{Object} &
\colhead{UT} &
\colhead{Dataset}  &
\colhead{Exp. time} }
\startdata  
RX\,J2329 & 2002-11-28 17:20:11 & o6li06010 & 730s \\
BZ\,UMa   & 2002-10-25 04:37:52 & o6li24010 & 700s \\ 
EY\,Cyg   & 2003-02-10 09:06:48 & o6li0v010 & 700s \\ 
CE315     & 2003-04-10 19:14:41 & o6li05010 & 900s \\
\enddata  
\end{deluxetable}  

\begin{deluxetable}{rcccc}
\tablecolumns{5}  
\tablewidth{0pc}  
\tablecaption{\label{t-linefluxes}Emission line fluxes}
\tablehead{  
\colhead{Line} &
\colhead{RX\,J2329} &
\colhead{CE315} &
\colhead{BZ\,UMa}  &
\colhead{EY\,Cyg}  \\
&
\colhead{$10^{-15}$\,\ecs} &
\colhead{$10^{-15}$\,\ecs} &
\colhead{$10^{-14}$\,\ecs} &
\colhead{$10^{-15}$\,\ecs} }
\startdata  
\Line{N}{V}{1240}   & $45\pm5$ & $58\pm4$ & $63\pm3$ & $57\pm5$ \\
\Line{Si}{IV}{1400} & $35\pm5$ & $<3$     & $25\pm3$ & $21\pm2$ \\
\Line{C}{IV}{1550}  & $<5$     & $\mathbf{<4}$ & $\mathbf{<4}$ & $\mathbf{<4}$     \\
\Line{He}{II}{1640} & $17\pm6$ & $20\pm3$ & $11\pm4$ & $10\pm3$ \\
\enddata  
\end{deluxetable}  

\newpage

\end{document}